\documentclass[showpacs,twocolumn,preprintnumbers,amsmath,amssymb,prl,epsf,superscriptaddress]{revtex4-1}

\usepackage{graphicx}% Include figure files
\usepackage{dcolumn}% Align table columns on decimal point
\usepackage{bm}% bold math
\usepackage{hyperref}% add hypertext capabilities
\usepackage{amsmath,amssymb}
\usepackage{physics}
\DeclareMathOperator{\sinc}{sinc}
\usepackage{mathrsfs}
\usepackage{mhchem}
\usepackage{siunitx}
\usepackage{soul, xcolor}
\usepackage{array, makecell}
\usepackage[normalem]{ulem}

\setstcolor{violet}

\begin{document}

%\preprint{APS/123-QED}

\title{Towards third-order parametric down-conversion in optical fibers}

\author{Andrea Cavanna}
\affiliation{Max Planck Institute for the Science of Light, StaudtStra{\ss}e 2, 91058 Erlangen, Germany}
\affiliation{University of Erlangen-N\"urnberg, Staudtstra{\ss}e 7/B2, 91058 Erlangen, Germany}
\author{Jonas Hammer}
\affiliation{Max Planck Institute for the Science of Light, StaudtStra{\ss}e 2, 91058 Erlangen, Germany}
\affiliation{University of Erlangen-N\"urnberg, Staudtstra{\ss}e 7/B2, 91058 Erlangen, Germany}
\author{Cameron Okoth}
\affiliation{Max Planck Institute for the Science of Light, StaudtStra{\ss}e 2, 91058 Erlangen, Germany}
\affiliation{University of Erlangen-N\"urnberg, Staudtstra{\ss}e 7/B2, 91058 Erlangen, Germany}
\author{Erasto Ortiz-Ricardo}
\affiliation{Instituto de Ciencias Nucleares, Universidad Nacional Aut√≥noma de M\'{e}xico, Apartado Postal 70-543, 04510 Distrito Federal, Mexico}
\author{Hector Cruz-Ramirez}
\affiliation{Instituto de Ciencias Nucleares, Universidad Nacional Aut√≥noma de M\'{e}xico, Apartado Postal 70-543, 04510 Distrito Federal, Mexico}
\author{Karina Garay-Palmett}
\affiliation{Departamento de \'Optica, Centro de Investigaci\'on Cient\'ifica y de Educaci\'on Superior de Ensenada, Apartado Postal 2732, BC 22860 Ensenada, M\'exico}
\author{Alfred B. U'Ren}
\affiliation{Instituto de Ciencias Nucleares, Universidad Nacional Autónoma de M\'{e}xico, Apartado Postal 70-543, 04510 Distrito Federal, Mexico}
\author{Michael H. Frosz}
\affiliation{Max Planck Institute for the Science of Light, StaudtStra{\ss}e 2, 91058 Erlangen, Germany}
\author{Xin Jiang}
\affiliation{Max Planck Institute for the Science of Light, StaudtStra{\ss}e 2, 91058 Erlangen, Germany}

\author{Nicolas Y. Joly}
\affiliation{University of Erlangen-N\"urnberg, Staudtstra{\ss}e 7/B2, 91058 Erlangen, Germany} 
\affiliation{Max Planck Institute for the Science of Light, StaudtStra{\ss}e 2, 91058 Erlangen, Germany}
\author{Maria V. Chekhova}
\affiliation{Max Planck Institute for the Science of Light, StaudtStra{\ss}e 2, 91058 Erlangen, Germany}
\affiliation{University of Erlangen-N\"urnberg, Staudtstra{\ss}e 7/B2, 91058 Erlangen, Germany}

\date{\today}% It is always \today, today,

\begin{abstract}
Optical fibers have been considered an optimal platform for third-order parametric down-conversion since they can potentially overcome the weak third-order nonlinearity by their long interaction length. Here we present, in the first part, a theoretical derivation for the conversion rate both in the case of spontaneous generation and in the presence of a seed beam. Then we review three types of optical fibers and we examine their properties in terms of conversion efficiency and practical feasibility.  
\end{abstract}

\maketitle

\maketitle

%\begin{abstract}
%\end{abstract}

\section{Introduction}

Optical fibers have proved to be extremely efficient platforms for generating non-classical states~\cite{Takesue2004,Rarity2005,Li2005,Fan2005}. Whilst
bulk materials and, more recently, waveguides are also good candidates
for generating single- or two-photon states, the size and, most
importantly, the length of both is limited technically. Optical fibers
overcome this problem, the only real constraints on their length being
given by the optical loss and the homogeneity.

Direct generation of photon triplet states via the cubic interaction has
been a long standing goal in the field of quantum optics dating as far
back as the 1980's~\cite{Braunstein1987,Elyutin1990,Felbinger1998,Douady2004,Chekhova2005,Gravier2008,Corona2011,Moebius2016,Cavanna2016,Gonzalez2018}. The interest to this process, known as
third-order parametric down-conversion (TOPDC), is driven by the fact
that such interaction leads to the direct generation of a non-Gaussian
state~\cite{Braunstein1987}. It realizes a three-mode squeezing operator that differs
greatly from the two-mode squeezing operator that generates Gaussian
squeezed states. Photon triplet
states generated directly through the cubic interaction were observed only at microwave frequencies \cite{chang2019observation}. The photon triplet states generated so far at optical frequencies were mediated by the second-order susceptibility~\cite{Huebel2010,Guerreiro2014}. In the absence of
post-selection they do not display any non-Gaussian features. 

In this work we estimate the efficiency of TOPDC in various fibers
that have previously been suggested as promising platforms. Because
the expected rates of three-photon emission are in most cases tiny, we
also consider the case where TOPDC is seeded at the frequency of one
of the three emitted photons. Seeding dramatically increases the rate
of two-photon emission in the two remaining modes. Although the output
state of these modes in this case is expected to be the same~\cite{Okoth2019} as for
the usual two-photon spontaneous parametric down-conversion (SPDC), seeding can be used to study the TOPDC spectral features, similar to the way stimulated emission tomography (SET)~\cite{Liscidini2013} is used
to characterize SPDC.

\section{Theory}

For a monochromatic pump, the rate of transitions from the vacuum state to non-degenerate three-photon state that occupies modes 1, 2 and 3 is given by the Fermi golden rule,
\begin{equation}
    \label{eq:fermigolden}
    \Gamma_{i\rightarrow f}=\frac{2\pi}{\hbar^2}
      \abs{\bra{1_1,1_2,1_3}\hat{H}\ket{0_1,0_2,0_3}}^2 \delta(\Delta \omega),
\end{equation}
where $\bra{1_1,1_2,1_3}$ is the final three-photon state,
$\ket{0_1,0_2,0_3}$ is the initial vacuum state, $\hat{H}$ is the
Hamiltonian, and $\Delta \omega= \omega_p-\omega_1-\omega_2-\omega_3$.
The subscript $1,2,3$ denotes the state with frequency
$\omega_{1,2,3}$ and propagation constant $\vec{\beta}_{1,2,3}$. Accounting for the possible transition to several sets of modes as opposed to a single discrete three-mode state, we integrate Eq. (\ref{eq:fermigolden}) over a set of wavevector intervals. The rate of transitions into this set of intervals is then
\begin{multline}
    \label{eq:fermigoldenfiber}
     dR= \frac{2\pi}{\hbar^2} \left(\dfrac{L_q}{2\pi}\right)^3 \cross \\ 
     \abs{\bra{1_1,1_2,1_3}\hat{H}\ket{0_1,0_2,0_3}}^2
     \delta(\Delta \omega) \,d\beta_1\,d\beta_2\,d\beta_3,
\end{multline}
where $L_q$ is the quantization length. Using the dipole approximation and the third-order non-linear response gives the Hamiltonian~\cite{Okoth2019}
\begin{equation}\label{eq:hamiltonian}
    \hat{H}=-24 \epsilon_0 \chi_{eff}^{(3)}\int_{V_{int}} d^3\vec{r}{E_p^{(+)}E_1^{(-)}E_2^{(-)}E_3^{(-)}+h.c.},
\end{equation}
where $\chi_{eff}^{(3)}$ is the effective cubic susceptibility, $V_{int}$ is the volume of the cubic interaction, $\epsilon_0$ is the vacuum permittivity. The fields $E_p^{(+)}, E_1^{(-)}, E_2^{(-)}$ and $E_3^{(-)}$ relate to the pump and modes $1,2$ and $3$, respectively, the +(-) denote their positive (negative) frequency components. Please note that here we ignore terms for cross and self phase modulation in the Hamiltonian and reintroduce them at a later stage as their effect on the phase matching is well documented~\cite{Agrawal2012}. 
We describe strong macroscopic fields such as the pump (p) and later the seed (s) classically, 
\begin{equation}\label{eq:classfield}
    E^{(+)}_{p,s} = \mathcal{E}_{p,s} \widetilde{F}_{p,s}(x,y) ~ e^{i \beta_{p,s} z },
\end{equation}
whilst the weak fields in modes $1, 2, 3$ we describe using the quantum field operators  
\begin{equation}\label{eq:qfield}
    \hat{E}^{(-)}_n = \widetilde{\mathcal{E}}^*_{n} \widetilde{F}^*_n(x,y) \hat{a}^\dagger_{n} ~ e^{-i \beta_n z}.
  \end{equation}
Here, $z$ is the propagation direction in the fiber, the subscript $n=p,1,2,3$ denotes the field mode with frequency $\omega_n$, propagation constant $\beta_n$, refractive index $n(\omega_n)$, and group velocity $v_g(\omega_n)$, $\hat{a}^\dagger_{n}$ is the photon creation operator. The normalized transverse field distribution $\widetilde{F}_n(x,y)=\dfrac{{F_n}(x,y)}{\sqrt{\int \abs{F_n(x,y)}^2 dx\,dy}}$, where $F_n(x,y)$ is the unnormalized transverse field distribution~\footnote{$F_n$ are vectors but we omit vector notation for simplicity}. The field amplitudes are given by 
\begin{eqnarray}
    \mathcal{E}_{p,s}= \sqrt{\dfrac{ P_{p,s}}{2 c \epsilon_0 n(\omega_{p,s})}},\label{pump}\\
    \widetilde{\mathcal{E}}_{n}= i \sqrt{\dfrac{ \hbar \omega_n v_g(\omega_n)}{2 \epsilon_0 n(\omega_n) c L_q}}, \label{quant}
\end{eqnarray}
where $P_{p,s}$ is the power and $c$ the speed of light.

\subsection{Unseeded TOPDC}

Substituting Eqs.~(\ref{eq:classfield},\ref{eq:qfield}) into the Hamiltonian (\ref{eq:hamiltonian}) and integrating over the volume gives
\begin{multline}\label{eq:ham2}
    \hat{H}=-24 \epsilon_0 \chi_{eff}^{(3)} \dfrac{\mathcal{E}_{p} \widetilde{\mathcal{E}}^*_{1} \widetilde{\mathcal{E}}^*_{2} \widetilde{\mathcal{E}}^*_{3}}{A_{eff}} f(\Delta \beta)  \hat{a}_{1}^{\dagger} \hat{a}_{2}^{\dagger} \hat{a}_{3}^{\dagger}+h.c.
\end{multline}
Here we have introduced the effective mode area as 
\begin{equation}
   A_{eff}=\left(\int \widetilde{F}_p(x,y) \widetilde{F}^{*}_1(x,y) \widetilde{F}^{*}_2(x,y) \widetilde{F}^{*}_3(x,y)~dxdy\right)^{-1},
\end{equation}
and the phase matching function as
\begin{equation}
    f(\Delta \beta) = L \sinc\left(\frac{\Delta \beta L}{2}\right) \exp \left(i\frac{\Delta \beta L}{2} \right),
\end{equation}
where $L$ is the fiber length and the propagation constant mismatch is given by
\begin{equation}
    \Delta \beta= \beta_p-\beta_1-\beta_2-\beta_3-\beta_{NL}.
    \label{eq:mismatch}
\end{equation}
Here we account for the cross and self phase modulation terms, which we dropped from the Hamiltonian, by reintroducing a non-linear momentum mismatch term $\beta_{NL}= [\gamma_p-2(\gamma_{p,1}+\gamma_{p,2}+\gamma_{p,3})]P_p$, where $\gamma_p$ is the nonlinear coefficient for the pump self-phase modulation,
\begin{equation}
    \gamma_p=\dfrac{3 \chi^{(3)} \omega_p}{4 \epsilon_0 c^2 n_p^2 A^{(p)}_{eff}},
\end{equation}
and $\gamma_{p,n}$ is the nonlinear coefficient for cross-phase modulation, 
\begin{equation}
    \gamma_{p,n}=\dfrac{3 \chi^{(3)} \omega_n}{4 \epsilon_0 c^2 n_p n_n A^{(p,n)}_{eff}}.
\end{equation}
The effective mode area for the pump self phase modulation is 
\begin{equation}
    A^{(p)}_{eff}=\left(\int \abs{\widetilde{F}_p(x,y)}^4  ~dx ~dy \right)^{-1},
\end{equation}
whereas for the cross phase modulation term the effective mode area between the pump and the photon triplet is
\begin{equation} % Jonas: there should be ^3 for triplet modes and ^1 for pump, no?
    A^{(p,n)}_{eff}=\left(\int \abs{\widetilde{F}_p(x,y)}^2 \abs{\widetilde{F}_n(x,y)}^2 ~dx ~dy \right)^{-1}.
\end{equation} 
It is worth noting that if the peak pump power is low, for example if one works in the continuous-wave (CW) regime, then the cross and self modulation terms are negligible. 
Substituting Eq.~(\ref{eq:ham2}) into Eq.~(\ref{eq:fermigoldenfiber}) gives the following differential rate of triplet emission:
\begin{multline}\label{eq:diffratebeta}
    dR=\dfrac{\hbar}{\pi^2} P_p \gamma^{2}_{1,2,3}  \dfrac{\omega_1 \omega_2 \omega_3}{\omega_p^{2}} \cross \\ v_g(\omega_1) v_g(\omega_2) v_g(\omega_3)   \abs{f(\Delta \beta)}^{2} \delta(\Delta \omega) \,d\beta_1\,d\beta_2\,d\beta_3,
\end{multline}
where the nonlinear coefficient is 
\begin{equation}
    \gamma^{2}_{1,2,3}=\dfrac{9 ~ [\chi_{eff}^{(3)}]^2 \omega_p^{2}}{ \epsilon_0^2 c^4 n_p n_1 n_2 n_3 A^{2}_{eff}}.
\end{equation}
Eq.~(\ref{eq:diffratebeta}) can be rewritten in terms of frequency as opposed to propagation constant, using the relation for group velocity $\dfrac{d\beta_n}{d\omega_n}=\dfrac{1}{v_g(\omega_n)}$, as 
\begin{multline}\label{eq:diffratespontaneous}
    dR(\omega_1, \omega_2, \omega_3)=\dfrac{\hbar }{\pi^2} P_p \gamma^{2}_{1,2,3} \cross \\ \dfrac{\omega_1 \omega_2 \omega_3}{\omega_p^{2}}  \abs{f(\Delta \beta)}^{2} \delta(\Delta \omega) \,d\omega_1\,d\omega_2\,d\omega_3.
\end{multline}

Because the rate of unseeded TOPDC scales linearly with the pump
power, CW regime is more favourable in this case. Working in the pulsed regime would lead to competing non-linear processes that scale non-linearly with the pump power. Light from such processes could saturate detectors or interfere with the spontaneous generation of triplets.  

\subsection{Seeded TOPDC}

Stimulation of TOPDC requires a seed beam in one of the triplet modes, 1, 2 or 3. In all the cases below we choose to replace mode 3 with the seed, which we will denote by the subscript s. If the seed beam is strong then it can be described by a classical field, hence the Hamiltonian in Eq.~(\ref{eq:ham2}) can be rewritten as
\begin{multline}\label{eq:hamseed}
    \hat{H}_s=-24 \epsilon_0 \chi_{eff}^{(3)} \dfrac{\mathcal{E}_{p} \mathcal{E}^*_{s} \widetilde{\mathcal{E}}^*_{1} \widetilde{\mathcal{E}}^*_{2}}{A_{eff}} f(\Delta \beta)  \hat{a}_{1}^{\dagger} \hat{a}_{2}^{\dagger} +c.c.
\end{multline}
Here, the classical amplitude $\mathcal{E}^*_{s}$ of the seed is defined by Eq. (\ref{pump}) with $\omega_{s}=\omega_3$. From this it is clear that when a seed is present, the three-photon state is lost and instead a two-photon state is generated~\cite{Okoth2019}. Since we now consider the product of two classical fields there is an advantage in using a pulsed source. If the seed and pump pulses are overlapped spatially and temporally, then their product averaged over time will introduce an enhancement factor equivalent to the inverse duty cycle of the laser. 

In the pulsed regime the monochromatic approximation breaks down and the Fermi golden rule must be rewritten. Assuming a square pulse, the probability of a two-photon transition occurring over the pulse duration $t$ is
\begin{equation}
    \label{eq:fermigolden_s}
    \Pi_{i\rightarrow f} (t)= \frac{1}{\hbar^2}  
    \abs{\bra{1_1,1_2}\hat{H}_s\ket{0_1,0_2}}^2 \rho(\Delta \omega ,t).
\end{equation}
where
\begin{equation}
\rho(\Delta \omega,t)= t^2 \sinc\left(\frac{\Delta \omega t}{2}\right)^{2}.
\end{equation}
Again integrating over a set of wavevector intervals gives the differential number of transitions per pulse: 
\begin{multline}
    \label{eq:fermigolden_s}
    dN= \frac{L_q^2}{4\pi^2 \hbar^2}
    \abs{\bra{1_1,1_2}\hat{H}_s\ket{0_1,0_2}}^2 \rho(\Delta \omega,t)d\beta_1\,d\beta_2.
\end{multline}
Expanding out the quantum average gives
\begin{multline}\label{eq:diffseedk}
    dN=\dfrac{1}{\pi^2} P_p P_s \gamma^2_{1,2,s} \dfrac{\omega_1 \omega_2}{(\widetilde{\omega}_p)^2} \cross \\ 
		v_g(\omega_1) v_g(\omega_2) \abs{f(\Delta \beta)}^{2} \rho(\Delta \omega,t) \,d\beta_1\,d\beta_2,
\end{multline}
where the nonlinear interaction coefficient for the seeded process is
\begin{equation}
    \gamma^{2}_{1,2,s}=\dfrac{9 ~[\chi_{eff}^{(3)}]^2 (\widetilde{\omega}_p)^{2}}{\epsilon_0^2 c^4 n_p n_1 n_2 n_s A^{2}_{eff}}
\end{equation}
and $\widetilde{\omega}_p=\omega_p-\omega_s$.
Again, for convenience we rewrite Eq.~(\ref{eq:diffseedk}) in terms of frequency intervals, which yields
\begin{multline}\label{eq:diffseed}
    dN(\omega_1, \omega_2)=\dfrac{1}{\pi^2} P_p P_s \gamma^{2}_{1,2,s}\times \\
		\dfrac{\omega_1 \omega_2}{(\widetilde{\omega}_p)^{2}}  \abs{f(\Delta \beta)}^{2} \rho(\Delta \omega,t) \,d\omega_1\,d\omega_2.
\end{multline}
It is worth noting that $dN(\omega_1, \omega_2)$ and $dR(\omega_1, \omega_2, \omega_3)$ have different dimensionality, the former being the number of pairs emitted per pulse (dimensionless) and the latter being a rate of triplet emission (Hz) in the CW regime.

\section{Estimates and experimental evidence}
In order to ensure the efficient generation of photon triplets
  along the entire fiber length, phase matching has to be fulfilled,
see Eq.(\ref{eq:mismatch}). Normally this can be achieved by the
so-called intermodal phase matching, where different fields
  propagates in the fiber in different spatial modes. In the case of
triplet generation phase matching is usually found when the pump at
$\omega_p$ is in a high-order mode while the three photons are in the
fundamental mode. Coupling light into a high-order mode is quite
challenging and usually not very efficient. Furthermore, there are no
free parameters that allow one to tune the dispersion, therefore, for
a given optical fiber the phase matching frequencies are fixed. To
overcome these two problems we investigated different types of optical
fibers. The first one is a hybrid fiber that, due to the use of
  different guidance mechanisms, allows the phase matching between
  single-lobed modes. Not only does this simplifies the coupling, but it also increases the
  overlap between the pump and the generated fields. The second one
is a hollow-core fiber filled with xenon gas, which allows
tuneable phase matching by changing the gas pressure. Finally,
we consider a tapered fiber, which has a high overlap between
the two phase matching modes due to their high confinement and whose
dispersion can be tuned by changing the gas pressure of the 
  environment~\cite{Hammer2018}.

\subsection{Hybrid fiber}
\begin{figure}[htb]
\includegraphics[width=\columnwidth]{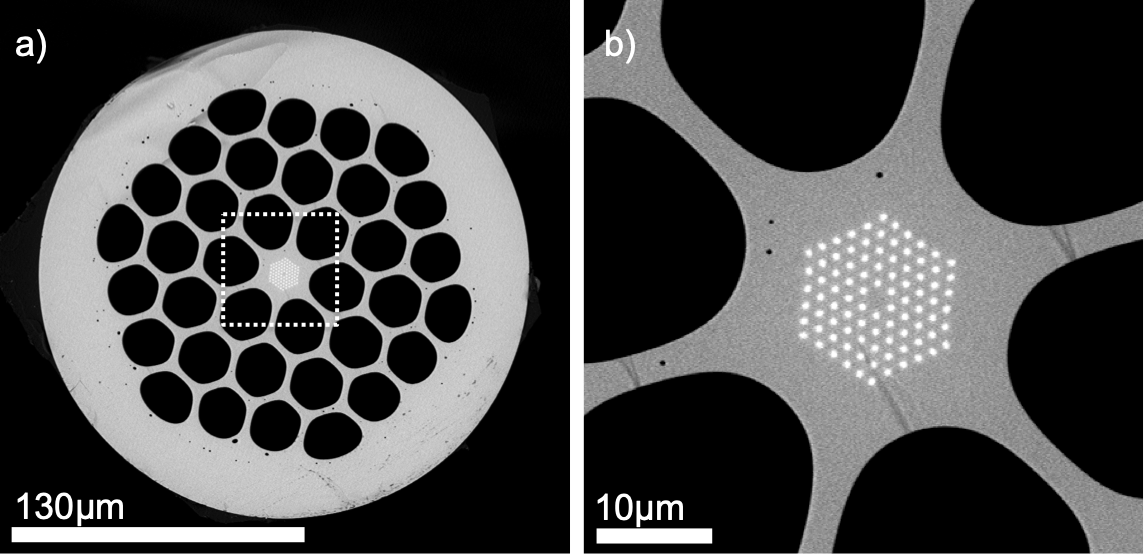}
\caption{SEM image of the hybrid fiber a) and its inner core structure b).}
\label{fig:structure_hybrid}
\end{figure}

\begin{figure}[htb]
\includegraphics[width=\columnwidth]{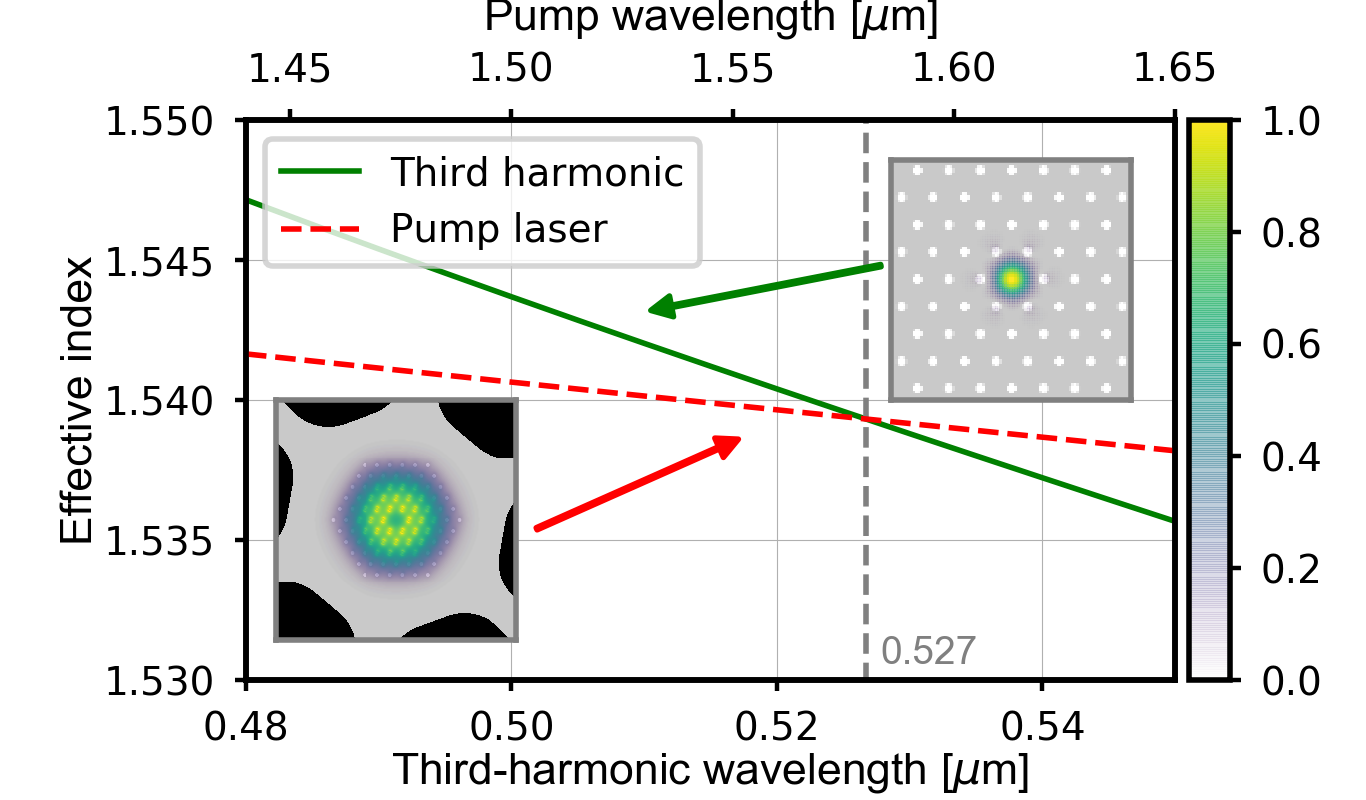}
\caption{Dispersion curves of the photonic bandgap mode and the infrared fundamental mode of the hybrid fiber. The insets show the calculated spatial distributions of the modes.}
\label{fig:phaseMatching_hybrid}
\end{figure}

The first fiber that we present is a solid core microstructured fiber
with a double core structure~\cite{Cavanna2016}. This fiber was designed to
circumvent the problem of coupling the pump beam into a high-order
mode by compensating the phase mismatch using modes of different size,
ideally both fundamental. This comes from the fact that the
propagation constant of a mode at a fixed wavelength can be decreased
by just reducing the mode diameter. The fiber has a photonic
bandgap (PBG) structure that guides the visible mode, surrounded by
hollow channels. The infrared mode is guided by total internal
reflection, the entire PBG structure acts as the core and the surrounding glass, together with the hollow channels, as the cladding. The inner PBG structure contains rods
made of high refractive index lead-silicate glass (Schott SF6)
embedded in low refractive index glass (Schott LLF1), see Fig.
\ref{fig:structure_hybrid} for the scanning electron micrograph (SEM)
of the fiber. The central rod of SF6 glass is replaced by one of LLF1
in order to create a defect that corresponds to the core. By carefully
tuning the rods diameter and the distance between them (the
  pitch), it is possible to create a bandgap that confines light at a
particular frequency~\cite{Birks2006}. In our case, the target wavelength
is 532 nm. For our fiber we choose a diameter of about 380 nm for the
SF6 glass rods and a pitch of 1.05 $\mu$m. The rods are arranged in
a hexagonal geometry with 5 concentric rings. On the one hand,
increasing the number of rings will decrease the guidance losses of
the visible mode but on the other hand, the dimensions of the PBG
structure define the infrared mode size and therefore its dispersion.
Fig.~\ref{fig:phaseMatching_hybrid} shows the dispersion and the intensity distributions of guided modes simulated with the finite-element model (FEM)~\cite{Cavanna2016}.
%figure: SEM

\begin{figure}[htb]
\includegraphics[width=\columnwidth]{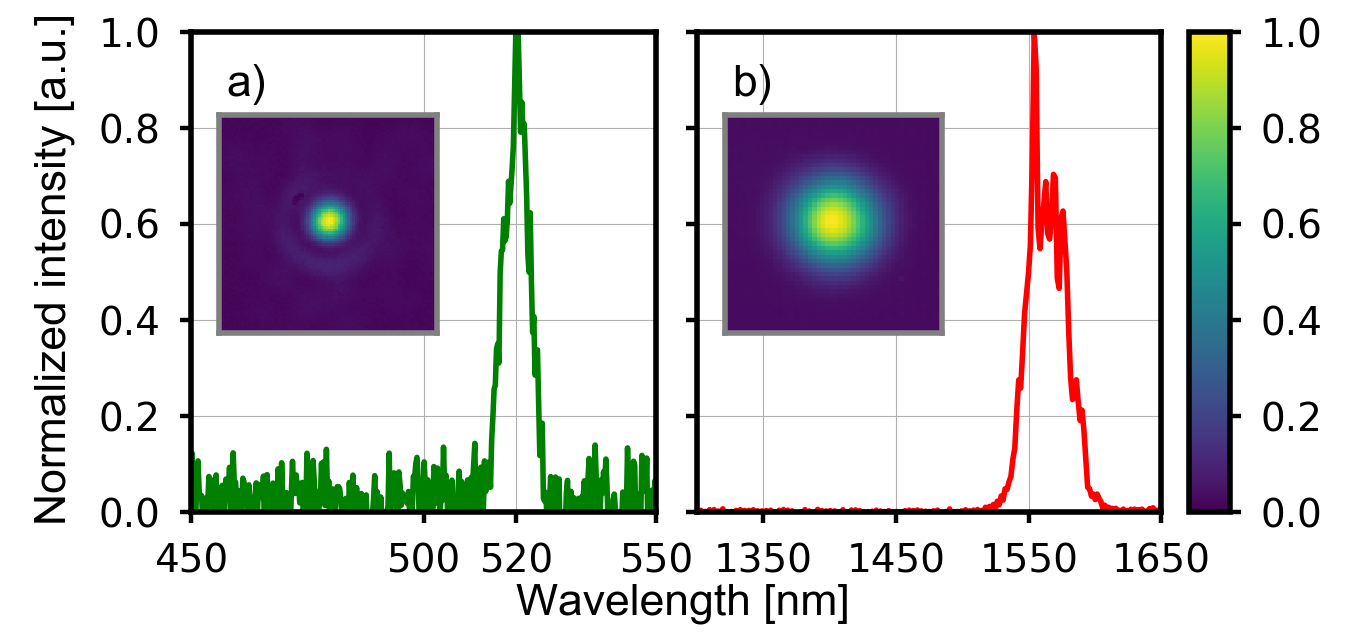}
\caption{Normalized spectra and near-field intensity distributions of the third harmonic a) and the pump b) for the hybrid fiber. The insets show the near field at the output of the fiber.}
\label{fig:THG_hybrid}
\end{figure}

Photon triplet generation can be seen as the reverse process of third harmonic generation (THG), with which it shares, among many features, the phase matching conditions. We therefore used the THG to test the phase matching for the manufactured fiber. Due to the small dimensions it is difficult to precisely control the fiber parameters and achieve phase matching at exactly $532$ nm. Therefore we used an optical parametric generator (OPG) pumped with the second harmonic of a Nd:YAG laser ($532$~nm) with $18$~ps pulse duration and $1$ kHz repetition rate~\cite{Cavanna2016a}. The output of the OPG was tuneable over more than $200$ nm with about 10 mW average power. Using this source we generated third-harmonic radiation in the optical fiber. The phase matching wavelength was found at $520$~nm and the near field intensity distribution of the mode was measured, see Fig.~\ref{fig:THG_hybrid}.

Starting from the mode electric fields, obtained with the FEM simulations, we calculated an effective area of $A^{(p,n)}_{eff}=218\,\mu$m$^2$. We considered the effective third-order susceptibility of SF6 glass  $\chi^{(3)}=1.15\cdot 10^{-21}$m$^{2}$V$^{-2}$~\cite{boling1978empirical}. Using Eq.~(\ref{eq:diffratespontaneous}) and assuming a pump power of 100 mW (CW), a length of $10$ cm, and a detection bandwidth of $150$ nm, we obtained an expected spontaneous triplet generation rate of $11$ Hz (Table.~\ref{table:spont}). 

\subsection{Gas-filled hollow-core fiber}
\begin{figure}[htb]
\includegraphics[width=\columnwidth]{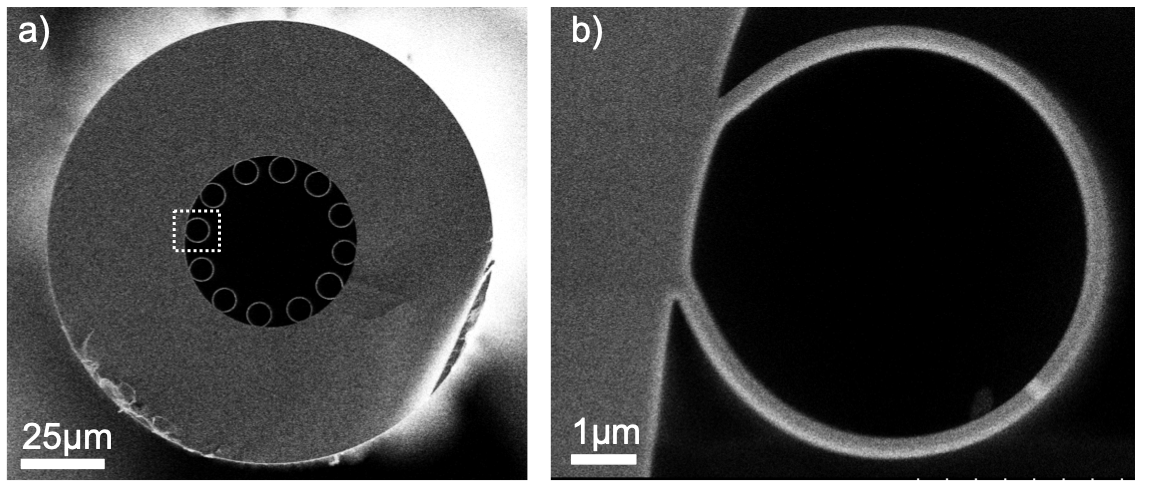}
\caption{SEM picture of the hollow-core structure a) with a detailed view of a capillary b)}
\label{fig:structure_singleRing}
\end{figure}

%structure and guidance mechanism
The second fiber that we present is a gas-filled hollow-core photonic-crystal fiber (PCF). Hollow-core fibers have the great advantage to allow tuneable phase matching by changing the pressure of the filling gas. For our experiment we choose to use Xenon due to its high nonlinearity \cite{bree2010method} and the very low reactivity. There are several geometries of hollow-core fibers, the one that we consider in this paper is known as single-ring fiber~\cite{pryamikov2011demonstration,wei2017negative}. It consists of a ring of capillaries attached to the inner surface of a glass tube, see Fig.~\ref{fig:structure_singleRing} a). 
Light is confined in the core region by means of an anti-resonant reflection at the core-cladding interface. The wall-thickness of the glass-capillaries defines the resonant frequencies at which the light is not guided. 
The dimensions of the capillaries determine the modal guidance of the fiber. The light is guided in the core only if there is no coupling with the capillary modes~\cite{uebel2016broadband}. 
For triplet generation we require a minimum guidance losses for
infrared light around $1600$ nm in the fundamental mode and at $532$ nm
for the phase matched high-order mode.  Depending on the core
diameter, the phase matching can be achieved at different gas
pressures~\cite{Nold2010}. A smaller core size usually implies higher guidance losses but at the same time also a higher gas pressure at which phase matching is achieved and therefore higher nonlinearity. The designed fiber has a core diameter of $38.7\,\mu$m surrounded by a ring of 12 capillaries, each with a diameter of $7.1\,\mu$m. For this configuration there is no coupling between the visible high-order mode and the capillaries' modes at the same frequency. 

\begin{figure}[htb]
\includegraphics[width=\columnwidth]{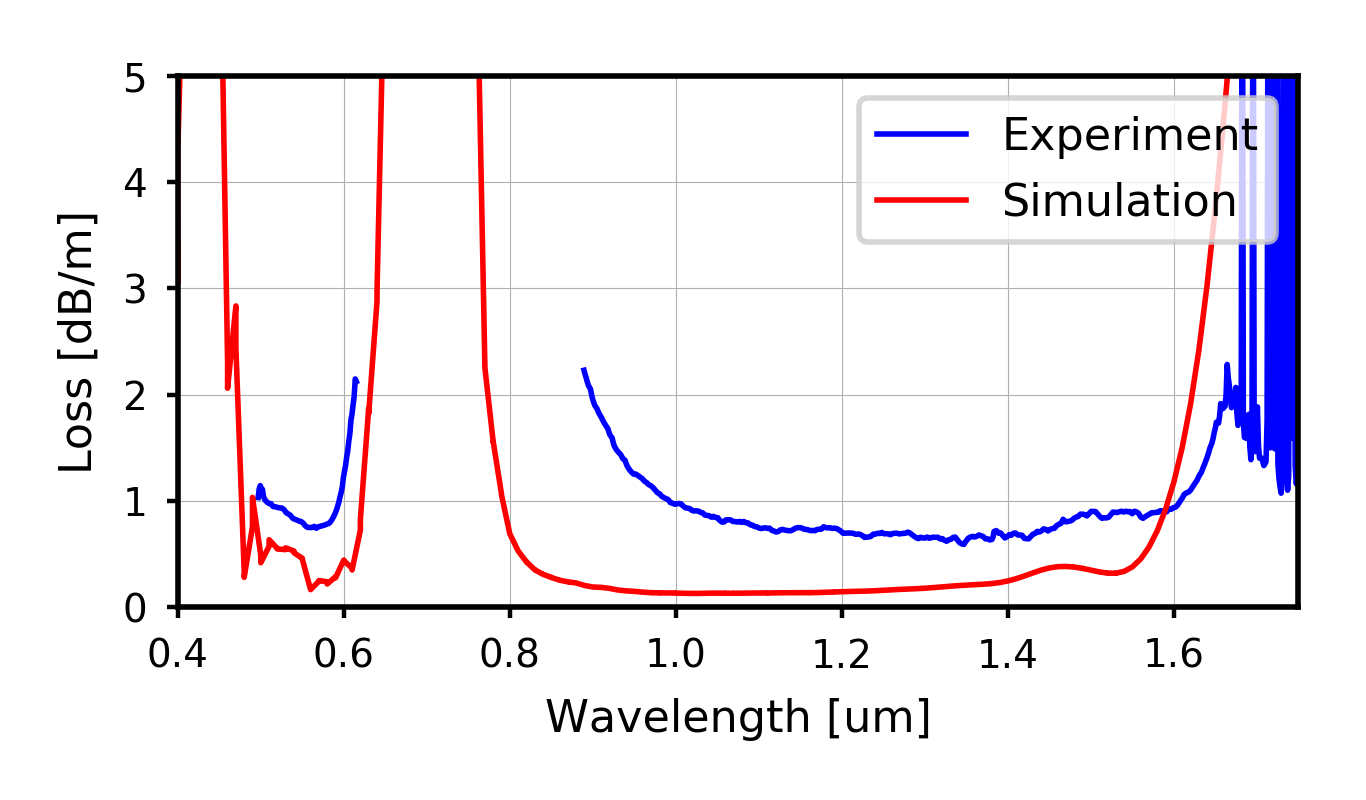}
\caption{The calculated (red line) and measured (blue line) loss for the single-ring PCF designed for the photon triplet generation.}
\label{fig:losses_singleRing}
\end{figure}

Using the formula given in Refs.~\cite{archambault1993loss,zeisberger2017analytic}, 
it is possible to estimate the resonance frequencies for  given parameters of the fiber. The glass thickness of the capillaries was chosen to be $350$~nm in order to set the resonant frequencies far from the region of interest.  Indeed, the drawn fiber, with corresponding glass thickness, has relatively low losses both at $532$ nm and around $1596$~nm. Figure~\ref{fig:losses_singleRing} shows that, in agreement with the calculation (red line), the measured (blue line) loss in the ranges of interest does not exceed $1$~dB/m.

\begin{figure}[htb]
\includegraphics[width=\columnwidth]{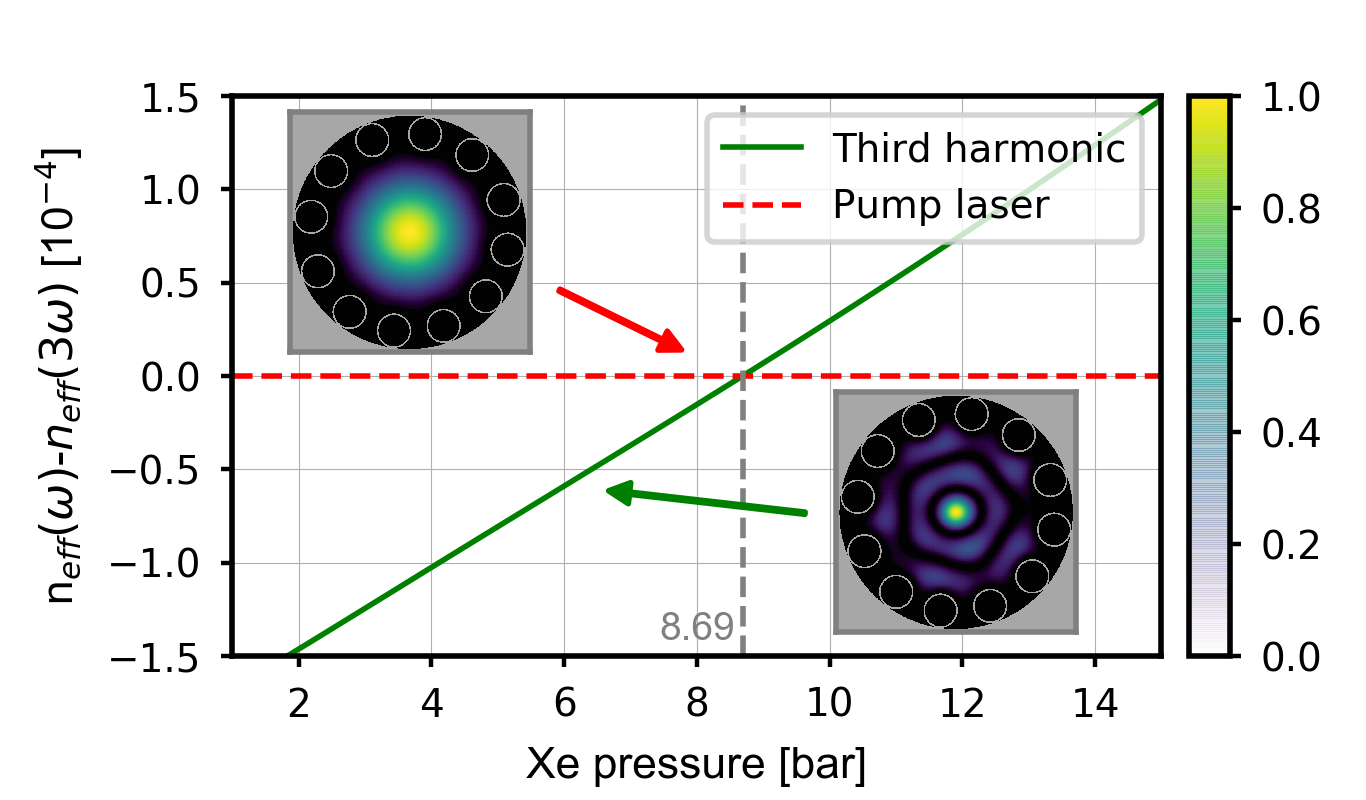}
\caption{Difference between the refractive indices  for 532 nm and 1596 nm versus the Xe pressure in a single-ring PCF. The insets show the simulated intensity distributions of the modes. The insets show the near field at the output of the fiber.}
\label{fig:phaseMatching_singleRing}
\end{figure}

Using a FEM model of the fiber we simulated the guided modes and their dispersion as a function of the Xe pressure, see Fig.~\ref{fig:phaseMatching_singleRing}. The phase matching was found at $8.7$ bar. 

\begin{figure}[htb]
\includegraphics[width=\columnwidth]{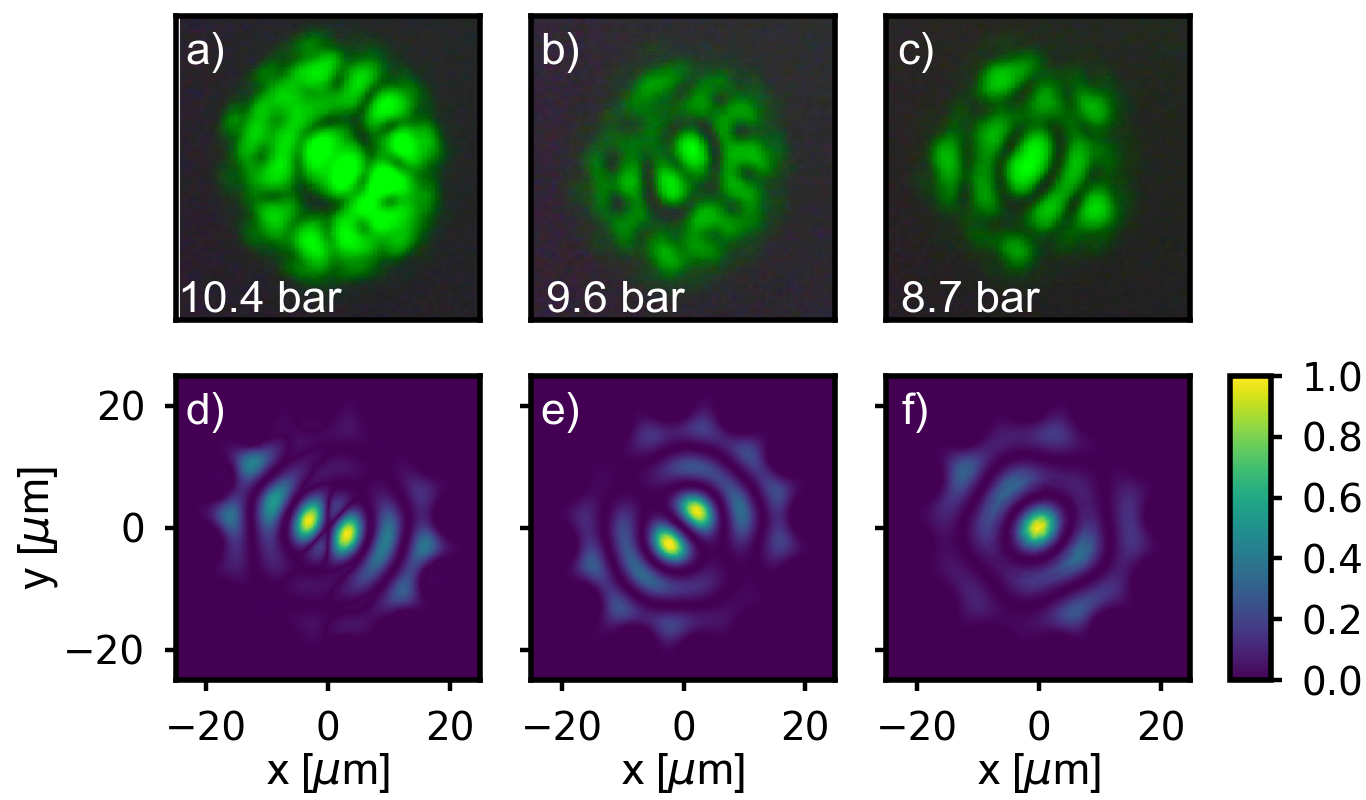}
\caption{Phase matched third-harmonic high-order modes generated by changing the Xenon pressure inside the gas chamber (from a) to c)). Simulated Poynting vector of the guided modes (from d) to f)).}
\label{fig:modes_singleRing}
\end{figure}

\begin{figure}[htb]
\includegraphics[width=\columnwidth]{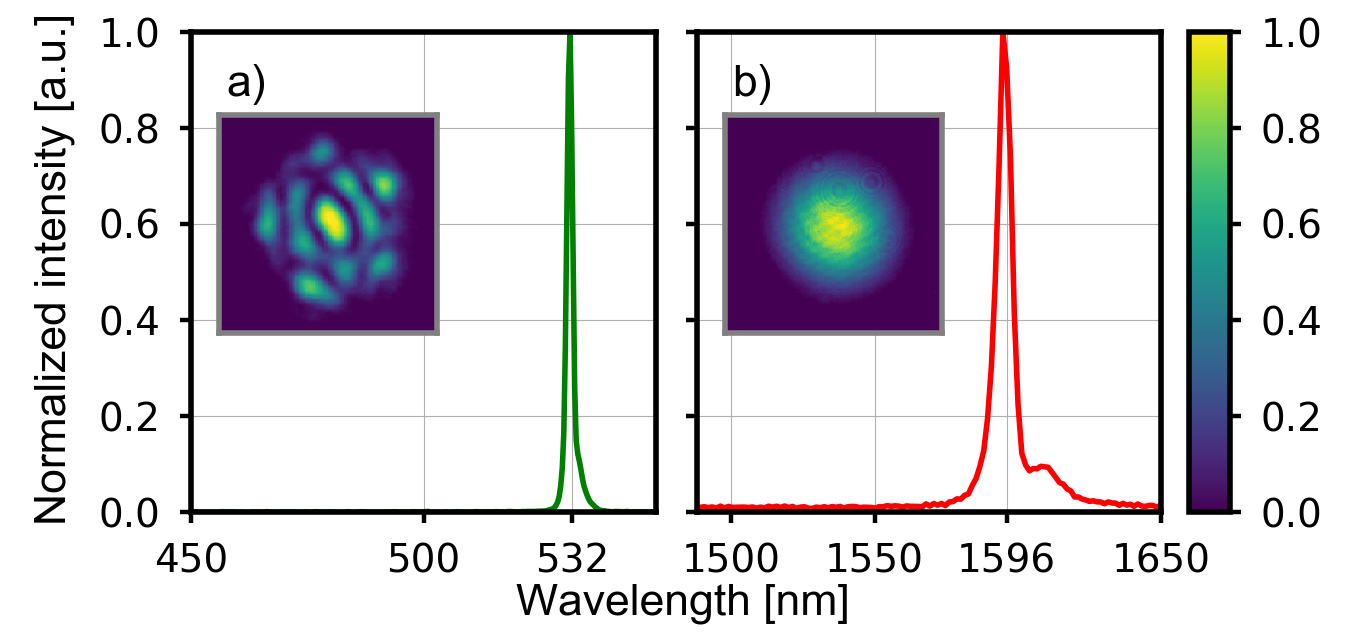}
\caption{Normalized spectra and measured near-field intensity distributions of the TH a) and the pump b) for Xe-filled PCF at 8.7 bar.}
\label{fig:THG_singleRing}
\end{figure}

In order to verify the phase matching we coupled the output of our OPG tuned at 1596 nm into the fiber core and observed the THG. To reduce the bandwidth of the pump beam we used a CW seed beam at 1596 nm in the OPG. The fiber was mounted in a gas-cell with a Xe pressure of about $15$ bar. By gradually reducing the gas pressure we achieved the THG for different high-order modes at $532$ nm, see Fig.~\ref{fig:modes_singleRing}. Our target mode, i.e. the one of the lowest possible order and therefore with the lowest effective area, was achieved at $8.7$ bar. Figure~\ref{fig:THG_singleRing} shows the recorded spectra and the intensity distributions for the targeted mode at $532$ nm and the fundamental IR mode.

The third-order nonlinearity of Xenon can be estimated from Refs.~\cite{bree2010method,lehmeier1985} $\chi^{(3)}_0=6.4\cdot 10^{-26}$m$^2$V$^{-2}$, where the subscript 0 indicates that the susceptibility was measured at a pressure of $1$~bar and at a temperature of $20$ degrees. The effective nonlinearity is proportional to the gas pressure, therefore in our system it is about 9 times larger than $\chi^{(3)}_0$. It is worth noticing that by reducing the core diameter and improving the fiber design (to reduce the confinement losses) not only the effective area will decrease but also the phase matching will move to higher pressure where the nonlinearity is higher. Due to the relatively low third-order nonlinearity of gases compared to solids, in
the hollow-core PCF the spontaneous triplet generation rate is only on the order of $10^{-5}$ Hz for pumping $1$ m of fiber with $200$ mW (Table.~\ref{table:spont}). Nevertheless, hollow-core PCF are very suitable for seeded generation, since they can guide very high power without being damaged. The only restrictions are the preparation and the coupling efficiency of the high-order pump mode. 

\subsection{Tapered fiber}
Consider now tapered optical fiber as a platform for
photon-triplet generation. Fiber tapers are manufactured by stretching a conventional step-index fiber (SMF28) over a scanning oxybutane flame~\cite{Birks1992}. Controlling the pulling speed and scanning range allows for a precise control over the taper parameters.
Decreasing the diameter of the fiber increases its effective nonlinearity, as well as drastically changes the waveguide contribution to the dispersion. The fiber tapers of interest here have a sub-micron waist diameter. At this diameter the initial fiber core is reduced so much that it does not play any significant role in the guiding mechanism, and the dispersion of the taper waist is given by the dispersion of a silica rod with the corresponding diameter in vacuum.~\cite{SnyderLove}
For triplet generation, the use of tapered optical fiber requires inter-modal phase matching similar to the case of gas-filled hollow-core fibers. 

\begin{figure}[!h]
  \centering
  \includegraphics[width=7.5cm]{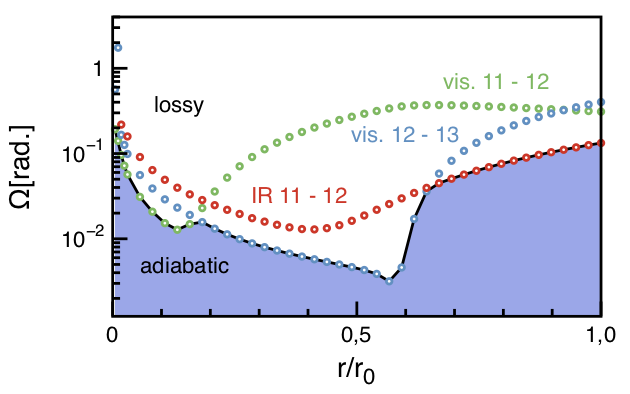}
  \caption{Maximal angle allowing an adiabatic transition for  HE$_{11}$ and HE$_{12}$, with the pump in the visible and the generated signal in the IR. The taper should be designed
    such that the local angle always lies within the shaded area to ensure adiabaticity. The red dots correspond to the maximum angle to avoid coupling from the IR HE$_{11}$ to the next relevant higher order mode HE$_{12}$. Green/Blue dots are the criterion for the visible light to avoid coupling between the HE$_12$ and the next relevant lower/higher order modes.}
  \label{fig:localAngle}
\end{figure}

As shown by Corona \emph{et al.}~\cite{Corona2011}, the most favourable case occurs when the visible pump light is guided in the HE$_{12}$ mode and the generated photon triplet-state is in the fundamental mode of the fiber. It is essential to ensure an adiabatic transition for both spatial modes involved in the generation process. Therefore, the transition profile of the taper must be designed very carefully. In particular, the local transition angle $\Omega(z)$ must remain small enough to avoid coupling between the fiber modes. The maximum angle allowed for adiabatic mode conversion between the untapered fiber and the taper waist is given by \cite{Love86}: 
\begin{equation}
  \label{eq:AdiabaticAngle}
  \Omega = \dfrac{\rho}{2\pi}\left( \beta_i  - \beta_{i\pm1} \right),
\end{equation}
where $\rho$ is the local core radius of the taper, $\beta_i(z)$ the
local wavenumber of the $i-th$ mode to couple into the waist taper and
$\beta_{i+1}$ the closest mode, to which coupling should be avoided.
Here, we used a finite-difference eigenmode solver to
compute the wavenumber for the HE$_{11}$ and HE$_{12}$ for the IR and
the same two modes plus the HE$_{13}$ in the visible
(Fig.~\ref{fig:localAngle}). In this calculation the
HE$_{12\text{,IR}}$ and HE$_{13\text{,vis}}$ are guided at the
cladding-air interface at all places along the transition, while the
other modes are guided in the core of the un-tapered fiber and evolve into cladding modes along the transition. Using
Eq.~\ref{eq:AdiabaticAngle} we can calculate the three different
curves in Fig.~\ref{fig:localAngle} corresponding to the adiabatic
angles avoiding coupling between HE$_{11}$ and HE$_{12}$ in both the
visible as well as the IR, as well as avoiding coupling between
HE$_{12}$ and HE$_{13}$ in the visible. The actual transition should
always lie below all three curves.

Pumping at $532$~nm, the diameter of the waist should be $790\ $nm for degenerate photon triplet generation. In this case, the  effective area $A_{eff} = 7.9\, \mu m^{2}$  and using $\chi^{(3)}_{eff} = 10^{-22}$ m$^2$V$^{-2}$ for silica we
can evaluate a generation rate as high as $3.2$ Hz for
a $10\ $cm-long taper with a pump of $20\ $mW (Table~\ref{table:spont}). %\JHadd{\it We have to relate fabrication tolerances to the spread of the JSI, generation rate is not really influenced by this, they will spread out too much to be practical}

It is important to note that the exact value of the diameter is extremely critical. With the same pumping condition but a tapered waist of $791$ nm the phase matching is no longer degenerate and the emission bandwidth is 130 nm (see Appendix). The same effect can be obtained by changing the pump wavelength by about $0.5$~nm. Such stringent fabrication tolerance cannot be met in practice. We can however circumvent this difficulty by encapsulating the tapered fiber inside a gas-cell filled with 
a gas under adjustable pressure~\cite{Hammer2018}.

%Figure~\ref{fig:influence_pressure} shows how the environmental
%pressure of {\AC Argon can modify the required phase-matching conditions.
%Using Argon up to 150 bar relaxes the fabrication tolerance by almost 40 nm. } % \sout{For Xenon the maximum pressure that we can use at ambient temperature is 55 bar. At higher pressure it becomes supercritical~[Azhar?] and it is technically more difficult to work with.}

%\begin{figure}[!h]
%  \centering
%  \includegraphics[width=\columnwidth]{PM_taper_Argon.png}
%  \caption{Influence of the environmental Ar-pressure to the
%    phase-matching conditions.}
%  \label{fig:influence_pressure}
%\end{figure}

\begin{figure}[!h]
  \centering
  \includegraphics[width=\columnwidth]{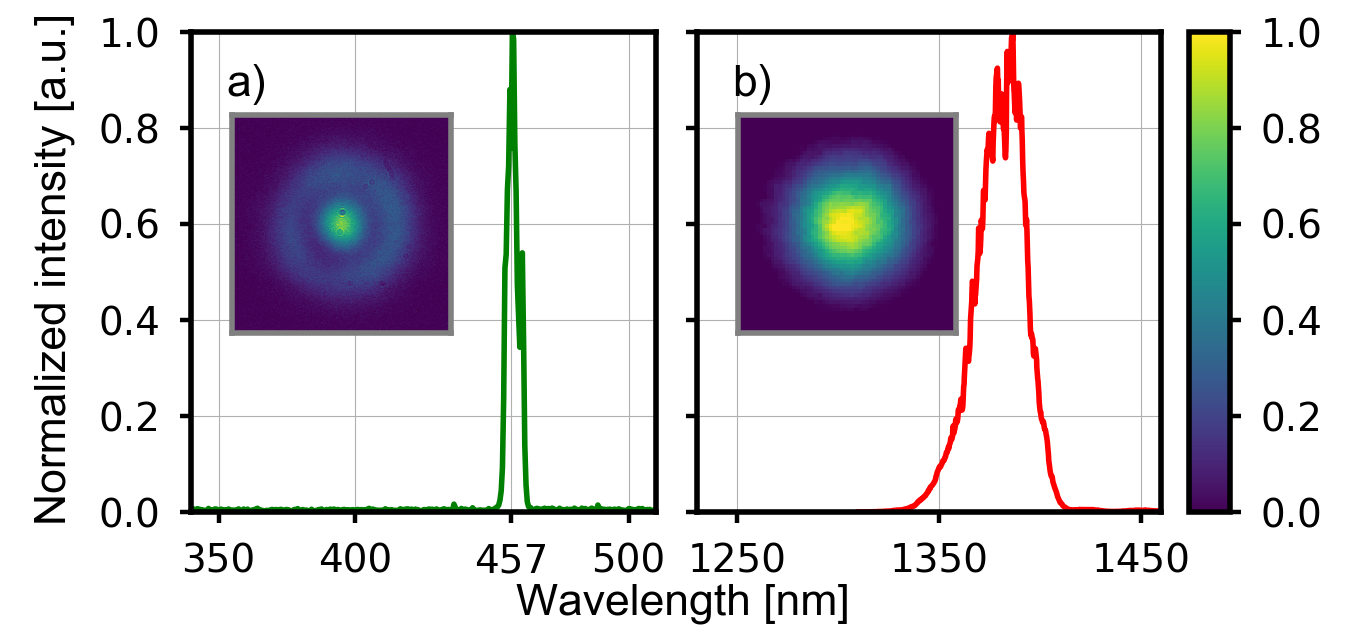}
  \caption{Normalized spectra and measured near-field intensity distributions of the TH a) and the pump b) for tapered single-mode fiber in vacuum. The insets show the Poynting vector distributions at the output of the fiber.}
  \label{fig:THG_taper}
\end{figure}

Similarly to the other fibers, third-harmonic generation was measured in the fiber taper. The fiber was pumped with a tunable laser having 0.25 MHz repetition rate, and 160 fs pulse duration and central wavelength 1375 nm, see Fig.~\ref{fig:THG_taper}. The third harmonic was generated at 457 nm in the expected HE$_{12}$ guided mode.   

\begin{table*}[htb]
\begin{tabular}{ ||c|c|c|c|c|c|c|c|| } 
\hline
\makecell{Fiber \\ Type} 
& \makecell{Pump \\ power ($P_{p}$)} 
& \makecell{$\chi^{(3)}_{eff}$ \\ ($10^{-22}  m^2V^{-2}$)}
& \makecell{Pump \\ Wavelength [$\mu \text{m}$]}
& \makecell{Effective Area \\ ($A_{eff}$)[$\mu \text{m}^{2}$]}
& \makecell{Length \\ ($L$) [cm]}
& \makecell{Detect. bandwidth \\ ($\Delta \lambda$) [nm]}
& \makecell{Triplet \\ Rate [Hz]} \\
\hline
 Hybrid core & 
 100 mW & 
 11.5 &
 0.526865 &
 218 &
 10 &
 150 &
 11 
 \\ 
\hline
 Hollow-core  & 
 200 mW & 
 0.043 &
 0.532 &
 19200 &
 100 &
 150 &
 $5.5\cdot10^{-6}$ 
 \\
\hline
 Tapered  & 
 20 mW & 
 2.5 &
 0.532 &
 7.89 &
 10 &
 150 &
 3.2 
 \\
\hline
\end{tabular}
\caption{Comparison of the expected triplet generation rate for different fiber types under realistic experimental conditions.}
\label{table:spont}
\end{table*}

\begin{figure}[!h]
  \centering
  \includegraphics[width=\columnwidth]{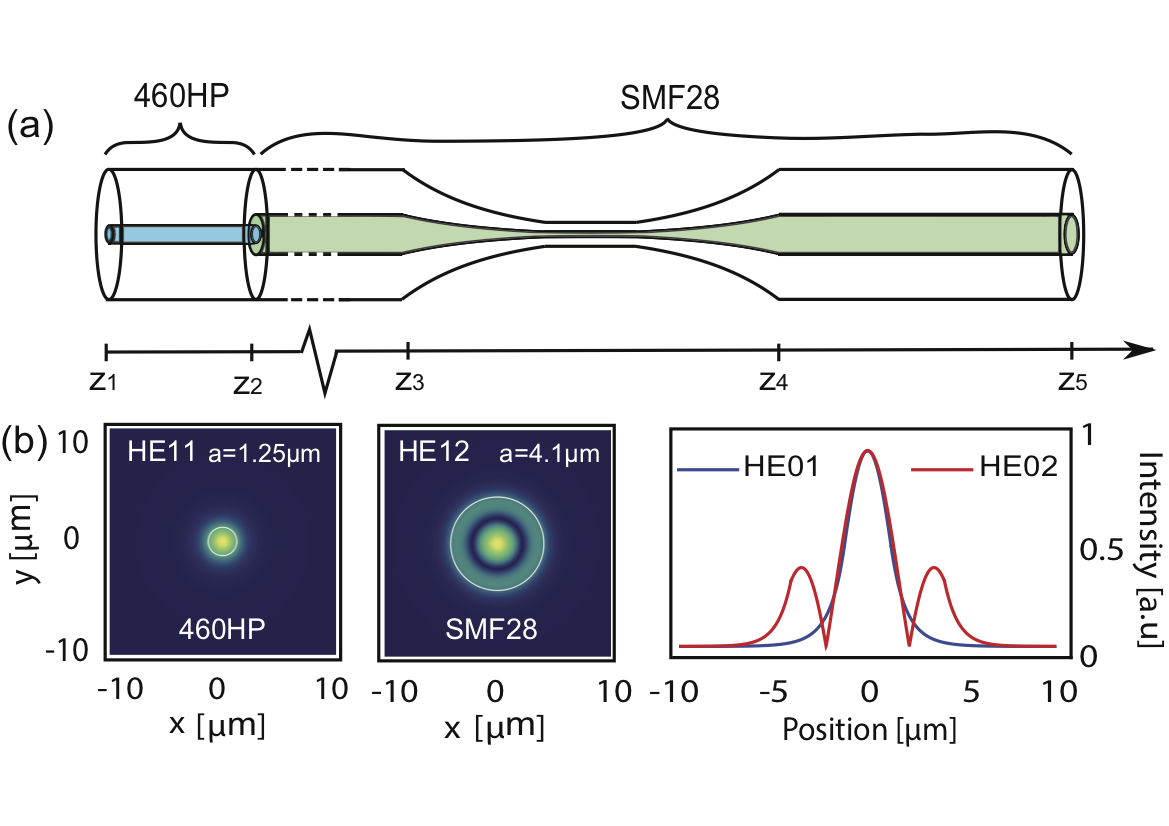}
  \caption{(a) Proposed device based on 460HP and SMF28 fibers spliced together, with a taper in the SMF28 stretch.  (b) HE$_{11}$ mode propagating in the 460HP fiber,  HE$_{12}$ mode propagating in the SMF28 fiber, and diametrical section of both modes.}
  \label{fig:overlapmodes}
\end{figure}

One of the challenges that we face when employing tapered fibers for TOPDC is the difficulty of coupling the pump at $532$ nm preferentially to the  HE$_{12}$ mode.  We have explored a solution which involves the use of two different types of fiber, 460HP and SMF28 spliced together, as shown schematically in Fig. \ref{fig:overlapmodes}(a), with the tapered region appearing along the SMF28 stretch of fiber. While the 460HP fiber is characterized by a core diameter of $2.5~\mu$m and is single-mode at $532$ nm, the SMF28 fiber with a core radius of $8.2~\mu$m is multi-mode at this wavelength (but single-mode at the TOPDC center wavelength of $1596$ nm).    

The concept that we exploit is that the fundamental mode HE$_{11}$ at $532$~nm propagating in the 460HP fiber excites preferentially,  upon arrival to the intra-fiber interface, the HE$_{12}$ mode 
 in the SMF28 fiber.  This occurs because: i) the mode field diameter of the HE$_{11}$ mode in the 460HP fiber is very similar to the diameter of \emph{the internal lobe} of the  HE$_{12}$ mode in the SMF28 fiber, and ii) the effective index of refraction of the incoming HE$_{11}$ mode  and that of the excited HE$_{12}$ mode are very similar, differing only by $\sim0.001$.  In Fig. \ref{fig:overlapmodes}(b) we plot the simulated incoming HE$_{11}$ mode, as well as the excited HE$_{12}$  mode, along with a diametrical section of these two modes overlapped in a single plot making it clear that the modes are remarkably similar within the internal lobe. This leads to an unusually high mode overlap between the incoming and excited modes (at approximately the same level as the overlap between the fundamental modes in the two fibers).  Note that the SMF28 can support higher-order modes but the overlap of these modes with HE$_{11}$ from the 460HP drops dramatically. 

\begin{figure}[!h]
  \centering
  \includegraphics[width=\columnwidth]{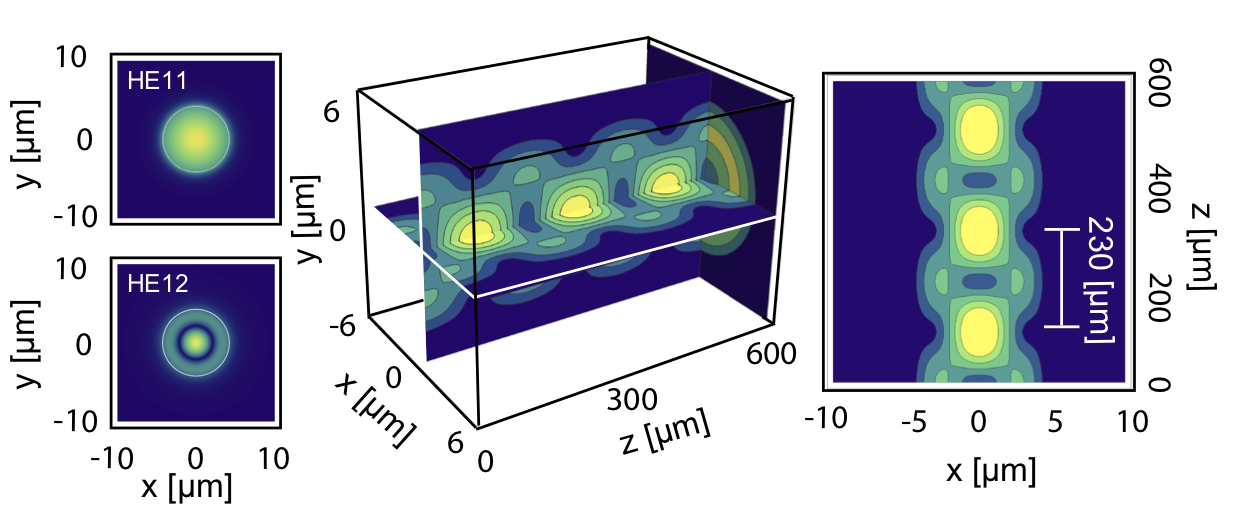}
  \caption{Left: HE$_{E1}$ and HE$_{12}$ modes in the SMF28 fiber. Middle: Resulting intensity  pattern in the $zy$ $xz$, and $xy$ planes. Right: Resulting intensity pattern in the $xz$ plane.}
  \label{fig:sum}
\end{figure}

%\JH{The presence in the SMF28 fiber of the modes HE$_{11}$, with transverse amplitude $E_1(x,y)$ and propagation constant $\beta_1$, and HE$_{12}$, with transverse amplitude $E_2(x,y)$ and propagation constant $\beta_2$,  leads to a total intensity $| E_1(x,y) + \exp(i [\beta_1-\beta_2]z ) E_2(x,y)|^2$ which exhibits spatial intensity oscillations with a period $2 \pi / (\beta_1-\beta_2) = 230\ \mu m$. }
In the SMF28, the presence of the HE$_{11}$ and HE$_{12}$ modes  (transverse amplitudes $E_1(x,y)$ and $E_2(x,y)$, and propagation constants $\beta_1$ and $\beta_2$ respectively) leads to a mode beating with a period $2 \pi / (\beta_1-\beta_2) = 230~\mu m$.
In Fig. \ref{fig:sum} we show the evolution of the intensity pattern along the fiber resulting from the beating of the fundamental HE$_{11}$ and higher-order mode HE$_{12}$.

\begin{figure}[!h]
  \centering
  \includegraphics[width=\columnwidth]{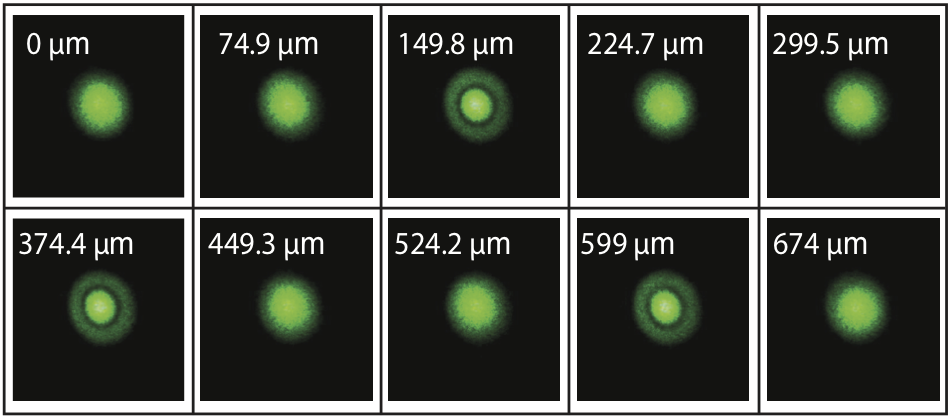}
  \caption{Imaged intensity pattern at the output of the fiber during taper fabrication, obtained from a video recorded by a CCD camera, with a $90$ms time between frames, corresponding to a $\sim75~\mu$m taper length differential between frames. }
  \label{fig:Video}
\end{figure}

We fabricated a device as described above, with 460HP and SMF28 fibers spliced together. The SMF28 is then tapered down to $\sim0.8~\mu$m diameter. As the taper is made its length increases. During this process it is then possible to observe, in real time, the evolution of the beating pattern between the two modes involved as a function of length by monitoring the output of the device. Fig.~\ref{fig:Video} shows the evolution of the intensity of the near field at the output of the SMF28 as a function of the added length. This is consistent with the simulated beating period of $230\ \mu$m, confirming the considered approach.

\begin{table*}[htb]
\begin{tabular}{ ||c|c|c|c|c|c|c||} 
\hline
\makecell{Fiber \\ type} 
& \makecell{Peak pump \\ power ($P_{p}$)} 
& \makecell{Seed \\ wavelength [$\mu \text{m}$]}
&\makecell{Peak seed \\ power ($P_{s}$)}
&\makecell{Inverse \\ duty cycle}
& \makecell{Detect. bandwidth \\ ($\Delta \lambda$) [nm]}
& \makecell{Number of \\ pairs (per pulse)} \\
\hline
 Hybrid core & 
 $5.0\cdot10^6$ W & 
 1.67 &
 $5.0\cdot10^5$ W &
 $5.0\cdot10^7$ &
 60 &
 $9.6\cdot10^6$ \\
\hline
 Hollow-core  & 
 $1.0\cdot10^7$ W & 
 1.605 &
 $5.0\cdot10^5$ W &
 $5.0\cdot10^7$ &
 200 &
 0.5 \\
\hline
 Tapered &
 $1.0\cdot10^6$ W & 
 1.6412 &
 $5.0\cdot10^5$ W &
 $5.0\cdot10^7$ &
 50 &
 $1.04\cdot10^8$ \\
\hline

\end{tabular}
\caption{Comparison of the expected pair generation rate for different fiber types in the case of a single seed and under realistic experimental conditions.}
\label{table:seed}
\end{table*}

\subsection{Seeding}

Although seeding third-order parametric down-conversion reduces the generated state from a three-photon to a two-photon one, it does dramatically increase the emission rate probability. Whilst the hybrid and tapered fibers have measurable spontaneous three-photon emission rates, due to the poor overlap and low cubic susceptibility the hollow-core fiber does not (see Table.~\ref{table:spont}). Hence to observe any effect in the hollow-core fiber, stimulating one of the three photons is a necessity. Despite loosing the three-photon correlations, seeding can still lead to information about the spontaneous three-photon state using stimulated emission tomography. Moving to the seeded emission probabilities and subsequently the seeded emission rates it is clear that one can observe photon pairs for each fiber type. Here we assume a pulsed laser with a $20$~ps pulse duration and 1 kHz repetition rate, hence working with single-photon detectors the number of photons detected for the hybrid and tapered fibers would be limited by the repetition rate. In order to perform a proper correlation measurement, the pump or the seed power has to be reduced to generate less then a pair per pulse. The estimated conversion rates with seeding are given in Table \ref{table:seed}.    

\section{Conclusion}
We presented three different types of fibers and estimated their potential viability as platforms on which to generate photon triplet states via a direct cubic interaction. We checked this for both the spontaneous and seeded regime.
The hybrid and tapered fibers proved to be the best fibers from which to observe photon triplets. However the former requires a tunable pump laser to satisfy phase matching due to its fixed geometry. Despite this drawback, the main advantage the hybrid fiber has over the other two designs is the ability to guide both the pump and triplets in a nearly Gaussian mode simultaneously. 
In contrast to the hybrid fiber, one can tune the phase matching in the the tapered fiber without changing the pump wavelength. This can be done by placing the tapered fiber into a gas under high tunable pressure. In addition, with the system we propose, it is possible to launch a high-order mode into the tapered fiber starting from a nearly Gaussian mode.   
Currently the design of the hollow-core fiber gives a very small conversion efficiency. However this can be improved by reducing the core size. By reducing the core diameter by only a factor of two, the conversion efficiency can be improved by around two orders of magnitude. %the number of triplets in the spontaneous regime will be 4.7e-4
In this case, a hollow-core fiber, with its high damage threshold and the pressure-dependent phase matching is a good candidate for seeding experiments.

\bibliographystyle{ieeetr}
\bibliography{bibliography_Tr_Fib}

\newpage
\section{\label{sec:Appendix} Appendix: Spectral density of three-photon emission}

The expression for $dR(\omega_1, \omega_2, \omega_3 )$ in Eq.\ref{eq:diffratespontaneous} can be simplified to a function of two variables by integrating $\omega_3$ over the delta function $\delta(\Delta \omega)$. Due to energy conservation this fixes the frequency of $\omega_3=\omega_p-\omega_1-\omega_2$. Thus we can rewrite Eq.\ref{eq:diffratespontaneous} as
\begin{multline}\label{eq:diffratespontaneous2}
    dR(\omega_1, \omega_2)=\dfrac{\hbar }{2\pi^2} P_p \gamma^{2}_{1,2,3} \cross \\ \dfrac{\omega_1 \omega_2 (\omega_p-\omega_1-\omega_2)}{\omega_p^{2}}  \abs{f(\Delta \beta)}^{2} \,d\omega_1\,d\omega_2.
\end{multline}
Plotting $S(\omega_1, \omega_2)=\dfrac{dR(\omega_1, \omega_2)}{\,d\omega_1\,d\omega_2}$ gives the spectral density of the three-photon emission.
\begin{figure*}
\includegraphics[width=18cm]{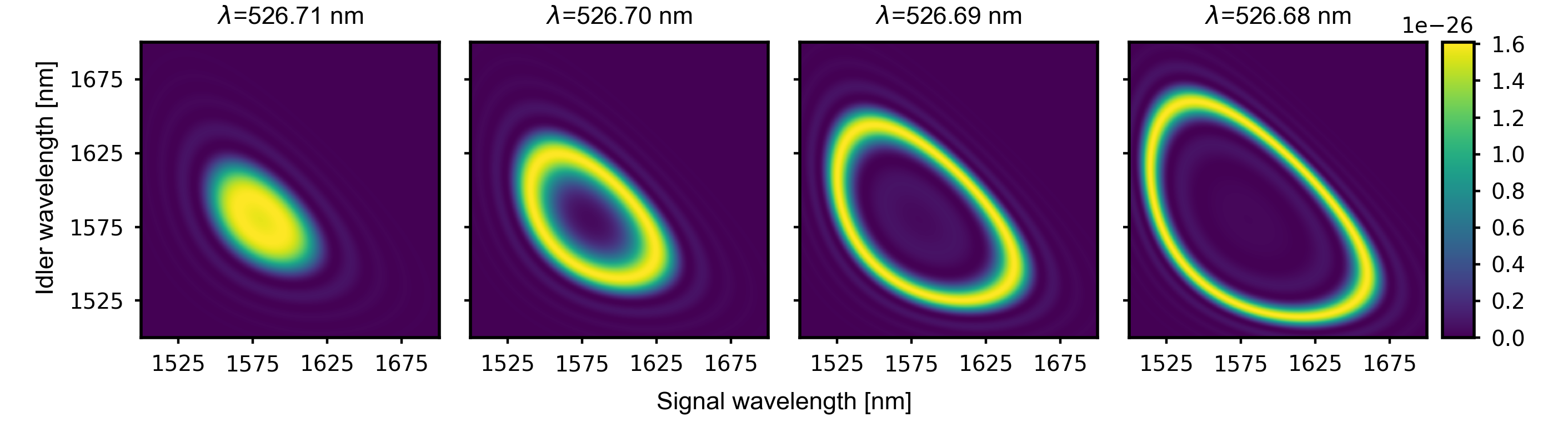} 
\caption{The spectral density $S(\omega_1, \omega_2)$ for the hybrid-core fiber at different pump frequencies.}
\label{fig:specdensHy}
\end{figure*}

Figure \ref{fig:specdensHy} ~shows the spectral density for the hybrid fiber for the parameters shown in Table \ref{table:spont}. Varying the pump wavelength by only 0.01 nm, one changes the spectral density vastly. This is due to the extremely different dispersion properties of the inner and outer core. In addition the spectral density is very broad, ranging over almost 150 nm. A combination of these two attributes makes it critical for the pump to be narrowband enough to minimize the collection bandwidth and increase the signal-to-noise ratio.

\begin{figure*}
\includegraphics[width=18cm]{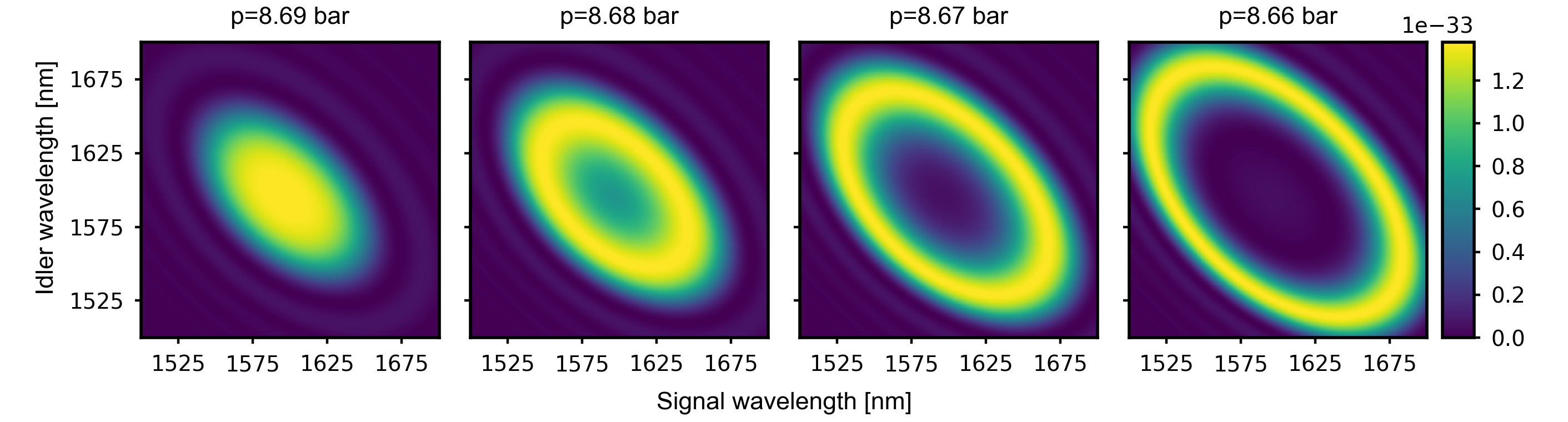} 
\caption{The spectral density $S(\omega_1, \omega_2)$ for the hollow-core fiber at different Xenon pressures.}
\label{fig:specdensHo}
\end{figure*}

Figure \ref{fig:specdensHo} shows the spectral density for the hollow-core fiber filled with Xenon gas and pumped at 532 nm. By reducing the gas pressure it is possible to pass from degenerate to non-degenerate phase matching.  Although the spectral width has a strong dependence on the pressure, this can be easily and precisely controlled with standard equipment.

\begin{figure*}
\includegraphics[width=18cm]{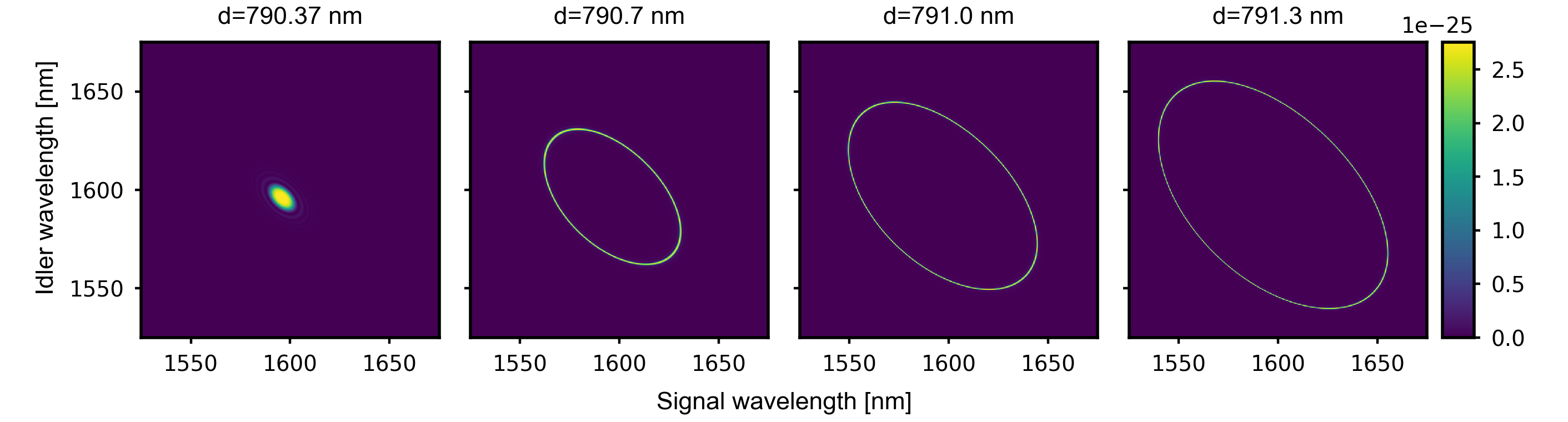} 
\caption{The spectral density $S(\omega_1, \omega_2)$ for the tapered fiber at different taper diameters.}
\label{fig:specdensTa}
\end{figure*}

Finally, in Fig.~\ref{fig:specdensTa} the spectral density for the tapered fiber is shown. The tapered region is highly dispersive due to the high confinement and increasing mode leakage into the ambient environment at longer wavelengths. This leads to a very narrow emission bandwidth. Although it means that the phase matching is critical, it is an advantage to work with the narrow spectrum as it means that the full emission spectrum can be collected.  
%\begin{figure*}
%\includegraphics[width=19cm]{specdensAll.pdf} 
%\caption{The spectral density $S(\omega_1, \omega_2)$ {\MC at different pump frequencies for the hybrid core fiber (top panels), the hollow core fiber (middle panels) and tapered fiber (bottom panels).}}
%\label{fig:specdensHy}
%\end{figure*}

%\section{\label{sec:Appendix} Appendix: Singly seeded Generation}

%{\CO Ideas: Seeding decreases bandwidth ,therefore reduces background and increases g(2), it can be used to reconstruct the three photon state via tomography, hude rates that should be easily measurable. Downsides: loss of three photon statistics and left with only two photon stats. Advantage is in using huge inverse duty cycle, this means competing effects. More problematic is it is easy to increase the number of photons to huge values, however if you are limited to single photon counting modules then the maximum count rate will be limited by the repetition rate of the laser. Therefore it is more important to look at the ratio between the pump photons and pair photons. This gives you the probability of detection and from this you can estimate how long you will have to wait to build up enough statistics to say whether we see something.}

\end{document}